\documentclass{natcomSI}
\usepackage{graphicx}
\usepackage{bm}
\usepackage{amssymb}
\usepackage{microtype}
\usepackage{xfrac}
\usepackage{makecell}
\usepackage{array}
\usepackage[version=4]{mhchem}
\usepackage{gensymb}
\usepackage[colorlinks=true]{hyperref}
\usepackage{xcolor}
\usepackage{caption}
\usepackage{soul}
\usepackage{ccaption}
\usepackage[symbol]{footmisc}
\usepackage[greek,english]{babel}

\usepackage{comment}

\newcommand{\gpi}{\textrm{\greektext p}}

\newcommand{\angstrom}{\text{\normalfont\AA}}

\title{Kramers Nodal Lines and Weyl Fermions in SmAlSi}

\author{Yichen Zhang${}^{1}\footnote[2]{These authors contributed equally: Yichen Zhang, Yuxiang Gao, Xuejian Gao}$  ,
Yuxiang Gao${}^{1}\footnotemark[2]$  ,
Xue-Jian Gao${}^{2}\footnotemark[2]$  ,
Shiming Lei${}^{2}$,
Zhuoliang Ni${}^{3}$,
Ji Seop Oh${}^{1,4}$,
Jianwei Huang${}^{1}$,
Ziqin Yue${}^{1,5}$,
Marta Zonno${}^{6}$,
Sergey Gorovikov${}^{6}$,
Makoto Hashimoto${}^{7}$,
Donghui Lu${}^{7}$,
Jonathan D. Denlinger${}^{8}$,
Robert J. Birgeneau${}^{4,9}$,
Junichiro Kono${}^{1,10,11}$,
Liang Wu${}^{3}$,
Kam Tuen Law${}^{2}\footnote[3]{The authors jointly supervised this work. Email: phlaw@ust.hk, emorosan@rice.edu, mingyi@rice.edu}$  ,
Emilia Morosan${}^{1}\footnotemark[3]$  ,
Ming Yi${}^{1}\footnotemark[3]$
}

\begin{document}

\maketitle

\begin{affiliations}
 \item Department of Physics and Astronomy, Rice University, Houston, Texas 77005, USA
 \item Department of Physics, Hong Kong University of Science and Technology, Clear Water Bay, Hong Kong, China
  \item Department of Physics and Astronomy, University of Pennsylvania, Philadelphia, PA 19104, USA
  \item Department of Physics, University of California, Berkeley, California 94720, USA
  \item Applied Physics Graduate Program, Smalley-Curl Institute, Rice University, Houston, Texas 77005, USA
  \item Canadian Light Source, Inc., 44 Innovation Boulevard, Saskatoon, SK S7N 2V3, Canada
 \item Stanford Synchrotron Radiation Lightsource, SLAC National Accelerator Laboratory, 2575 Sand Hill Road, Menlo Park, California 94025, USA
 \item Advanced Light Source, , Lawrence Berekeley National Laboratory, Berekeley, California 94720, USA
 \item Materials Science Division, Lawrence Berekeley National Laboratory, Berekeley, California 94720, USA
 \item Department of  Electrical and Computer Engineering, Rice University, Houston, Texas 77005, USA
\item Department of  Materials Science and Nanoengineering, Rice University, Houston, Texas 77005, USA
\end{affiliations}

\newpage

\begin{abstract}
\begin{center}
\section*{\large Abstract}
\end{center}
Kramers nodal lines (KNLs) have recently been proposed theoretically as a special type of Weyl line degeneracy connecting time-reversal invariant momenta. KNLs are robust to spin orbit coupling and are inherent to all non-centrosymmetric achiral crystal structures, leading to unusual spin, magneto-electric, and optical properties. However, their existence in in real quantum materials has not been experimentally established. Here we gather the experimental evidence pointing at the presence of KNLs in SmAlSi, a non-centrosymmetric metal that develops incommensurate spin density wave order at low temperature. Using angle-resolved photoemission spectroscopy, density functional theory calculations, and magneto-transport methods, we provide evidence suggesting the presence of KNLs, together with observing Weyl fermions under the broken inversion symmetry in the paramagnetic phase of SmAlSi. We discuss the nesting possibilities regarding the emergent magnetic orders in SmAlSi. Our results provide a solid basis of experimental observations for exploring correlated topology in SmAlSi

\end{abstract}

\newpage

\section*{Introduction}

Topology has emerged as a vibrant theme in modern condensed matter covering both gapped and gapless systems~\cite{Chiu2016,Hasan2010,Qi2011,Bernevig2013,Armitage2018}. Owing to the translational symmetry breaking and crystalline symmetry protection, solid crystals exhibit unique topological quasiparticles taking the forms of point~\cite{Armitage2018,Bradlyn2016,Tang2017,Wieder2016}, line~\cite{Bzdusek2016,Fang2016,Yang2018} and surface~\cite{Xiao2017,Wu2018,Kim2021,Wilde2021} degeneracies in the electronic band structure. Particularly interesting scenarios occur for noncentrosymmetric materials. Under broken inversion symmetry (IS), not only can Weyl fermions (WFs) arise, but also in all achiral space groups where the mirror or roto-inversion symmetry and time reversal symmetry (TRS) are preserved, crystals host doubly-degenerate lines known as the Kramers nodal lines (KNLs) that connect time reversal invariant momentum (TRIM) points~\cite{Xie2021,Hirschmann2021}. In contrast to nodal lines resulting from band inversion, the degeneracy of KNLs is robust against spin orbital coupling (SOC), and can only be lifted by explicitly breaking the mirror or roto-inversion symmetry. Therefore, when the KNLs cross the Fermi level, they produce touching points (spindle) or entire Fermi surfaces (octdong) that can be described by gapless Dirac fermions, leading to sizeable Berry curvature when gapped out by lifting the TRS, as well as quantized plateaus (thin film) or linear behavior (bulk limit) in optical conductivity for the latter case. When the mirror or roto-inversion symmetry is explicitly broken, the KNLs transform into Kramers Weyl points, which could lead to monopole-like electron spin texture, chiral magnetic effects, quantized circular photogalvanic effects~\cite{Chang2018_KWF}, or be utilized for constructing spin-orbit torque devices~\cite{He2021}. Furthermore, IS breaking could lead to effects on the superconductivity pairing mechanism~\cite{Bauer2012,Yip2014} and magnetic exchange interactions~\cite{Ghimire2013} such as parity mixing and Dzyaloshinskii-Moriya interaction. Therefore, the presence of WFs and KNLs in noncentrosymmetric materials offers an appealing playground for studying the interplay between topology and many-body physics such as unconventional superconductivity~\cite{Li2018,Shang2022,Bhattacharyya2019,Xu2019,Guan2016,Shang2022_KNL} and novel
magnetism~\cite{Ning2020,Zou2018,Duan2020,Wang2022_RKKY}.

Here we identify SmAlSi and its large isostructural family (LaPtSi-type) of materials to be candidates for hosting both KNLs and WFs~\cite{Chang2018}, existing in close proximity to the emergence of rich magnetic orders. WFs and surface Fermi arcs~\cite{Xu2017,Sanchez2020,Sakhya2022}, novel
magnetic and transport phenomena such as anomalous Hall and anomalous Nernst effect~\cite{Liu2021,Zhao2022,Meng2019,Destraz2020,Yang2021,Piva2021}, loop Hall effect~\cite{Yang2021,Piva2021}, singular magnetoresistance~\cite{Suzuki2019,Wang2022_1}, quantum oscillations~\cite{Wang2022_2,Xu2021,Cao2022}, and tunable magnetic domains~\cite{Sun2021,Xu2021_squid1,Xu2021_squid2} have recently been reported within this family. In addition, a Weyl-mediated Ruderman-Kittel-Kasuya-Yosida (RKKY) mechanism has been proposed in SmAlSi~\cite{Yao2022} and isostructural NdAlSi~\cite{Gaudet2021,Nikolic2021}. However, no theory and experimental efforts have been devoted to elucidating the novel topological feature of KNLs and examining the Fermi surface (FS) nesting possibilities. 
In our work, using angle-resolved photoemission spectroscopy (ARPES), density functional theory (DFT) calculations and quantum oscillation measurements, we provide evidence indicating this material family as viable candidates to host KNLs, in addition to numerous WFs appearing near the Fermi level, as observed in the paramagnetic state of SmAlSi. By a careful examination of the electronic structure near the Fermi level, we identify the wavevectors that could give rise to the potential nesting conditions in SmAlSi. Furthermore, a Lindhard susceptibility calculation is carried out to explore the origin of the incommensurate spin density wave (SDW). From a collective set of evidence including ARPES spectra on Fermi surface and at different temperatures, as well as first-principles calculations, it is suggested that the first SDW instability in SmAlSi is not likely to be purely driven by Weyl pocket nesting. 

\section*{Results}
\subsection{Structural, electronic, and magnetic properties}

\begin{figure}[t]
    \includegraphics[width=470pt]{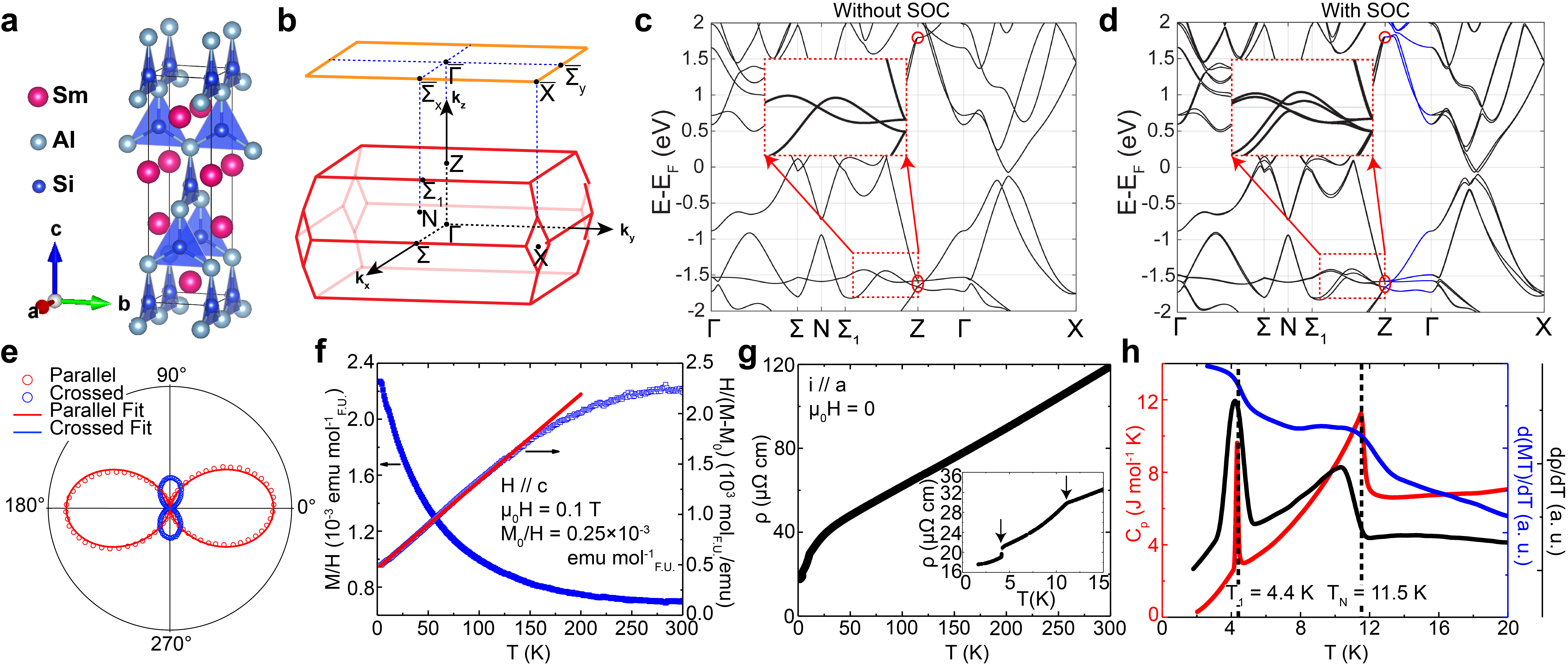}
    \centering
    \caption{\textbf{Crystal structure and electronic structure of SmAlSi.} \textbf{a} Crystal structure of SmAlSi where the blue triangles indicate the co-planar Al and Si atoms. \textbf{b} Bulk BZ (red) and the (001) surface projected BZ (orange). \textbf{c} and \textbf{d} Density functional theory (DFT) calculation of the bulk electronic band structure of SmAlSi without and with spin-orbit coupling (SOC), respectively. Kramers nodal lines along $\Gamma-Z$ are highlighted in blue and the 4-fold degeneracies at $Z$ are circled in red. The two pair of zoom-in boxes are to emphasize the effect of SOC splitting. The complete DFT plots can be found in Fig. S2. \textbf{e} Room temperature polarization-resolved optical second harmonic generation (SHG) measurements under 45 degree incidence on the (001) facet showing broken inversion symmetry. The parallel (red) and crossed (blue) curves are measured by keeping the incident and detecting polarizations parallel and perpendicular, respectively. \textbf{f} The magnetic susceptibility (filled blue symbols) measured at $\mu_0$\textbf{H} // $c~=~0.1$ T (left axis). The inverse susceptibility (open blue symbols) is shown with a Curie-Weiss fit (red line) (right axis). \textbf{g} Electrical resistivity as a function of temperature at \textbf{H} = 0. The inset shows a zoom-in view of the two transitions. \textbf{h} d$(MT)$/d$T$ (blue), d$\rho$/d$T$ (black), and specific heat $C_p$ (red) showing two transitions at T$_{\rm N}$ and T$_{\rm 1}$. }
    \label{fig:Basic}
\end{figure}

SmAlSi crystallizes in the LaPtSi structure type with space group $I4_1md$ (\#109), consistent with previous reports~\cite{Pikus2004,Wang2021} (See details of the powder X-ray diffraction in Fig. S1). The crystal structure and its three-dimensional (3D) Brillouin zone (BZ) are shown in Fig.~\ref{fig:Basic}a and b. Additionally, the (001) surface projected BZ (orange) is also illustrated in Fig.~\ref{fig:Basic}b, where the high-symmetry points are labeled with a bar on top to distinguish from those of the bulk. We note that all the ARPES, DFT and quantum oscillation results below are based on the nonmagnetic structure presented here. To firmly establish the IS breaking in the crystal structure, we carried out second harmonic generation (SHG) measurements. At room temperature, we observe strong SHG response, with the SHG polar plots shown in Fig.~\ref{fig:Basic}e. Further fits (See Supplementary Note 1for details) with the nonlinear susceptibility tensor associated with the point group $C_{4v}$ capture well the angle dependent patterns in the polar plots, confirming the absence of IS. 

To evaluate the topological electronic states in SmAlSi, we performed DFT calculations without and with SOC. The bulk band structures are shown in Fig.~\ref{fig:Basic}c and d. SOC generally lifts the band degeneracy except in the following two cases. First, the band degeneracy remains at the TRIM points due to the Kramers degeneracy~\cite{Chang2018_KWF}. Second, the band degeneracy can also be protected by achiral little group symmetry along certain lines that connect the TRIM points across the BZ, as suggested by recent theoretical work~\cite{Xie2021,Hirschmann2021}. These are the KNLs (Supplementary Table S1 and S2). Furthermore, as can be seen from Fig.~\ref{fig:Basic}d, an extra 4-fold degeneracy emerges at $Z$ due to the nonsymmorphic symmetry of the crystal, in contrast to the 2-fold degeneracy at $\Gamma$ (Supplementary Table S1 and S2). Besides the KNLs, a series of Weyl points also appear, and both of these will be discussed in detail in the next two sections. 

The magnetic properties of SmAlSi are relevant to the understanding of these topological states. We therefore characterize the magnetism in SmAlSi with magnetization, electrical resistivity and specific heat measurements. Figure ~\ref{fig:Basic}f-h shows the $\mu_{\rm 0}H$ = 0.1 T temperature-dependent magnetic susceptibility, $\mu_{\rm 0}H$ = 0 electrical resistivity and specific heat measurements, respectively. At low temperatures, they reveal two magnetic phase transitions, an antiferromagnetic transition at T$_{\rm N}$ = 11.5 K, and a spin reorientation transition at T$_{\rm 1}$ = 4.4 K, marked by the peaks in $C_p$ and the derivatives d$(MT)$/d$T$ and d$\rho$/d$T$. This is consistent with a recent report\cite{Yao2022}, while earlier work only observed the higher temperature transition at T$_{\rm N}$~\cite{Wang2021,Xu2021,Cao2022}. The high temperature inverse susceptibility (right axis, Fig. \ref{fig:Basic}f) shows Curie-Weiss behavior up to $\sim$ 150 K, and the corresponding linear fit (solid line) yields an effective moment $\mu_{\rm eff}^{\rm exp}$ = 0.96 $\mu_{\rm B}$, close to the predicted value for Sm$^{\rm 3+}$, and a negative Weiss temperature $\theta_{\rm W}$ = -53.2 K. In the following, we focus on the $\Gamma-Z$ high symmetry direction to provide evidence of KNLs in the paramagnetic state of SmAlSi through a combination of first-principles calculations, soft X-ray (SX) and vacuum ultraviolet (VUV) ARPES.

\subsection{Kramers nodal lines}

Kramers nodal lines, which are doubly degenerate lines connecting TRIM points, are predicted to exist in all nonmagnetic noncentrosymmetric achiral crystals in the presence of SOC~\cite{Xie2021}. The topological nature of KNLs is manifested in a quantized winding number $\gpi$, embodying a quantum version of Dirac solenoids. 
To date, KNLs remain a theoretical concept, with a number of proposed host materials. Here, we identify the family of compounds $R$Al(Si,Ge) ($R$ = La - Sm)
as viable hosts for KNLs.
Notably, SmAlSi possesses two types of nodal lines in the presence of SOC, KNLs and those protected by nonsymmorphic symmetry (See Fig. S2g). A complete list of nodal lines in SmAlSi is included in the Supplementary Note 2. Here, we focus on the identification of the KNLs, as illustrated in Fig.~\ref{fig:KNL}a. The KNLs themselves can be separated into two types: those that connect different TRIM points and those that connect the same TRIM points. The representative of the first type is a straight KNL connecting TRIM points $\Gamma$ and $Z$ (marked by thick red line), which is protected by the $C_{4v}$ little cogroup along $\Gamma-Z$. A representative of the second type, the curved KNLs connecting the $N$ points (blue curves lying within the $k_x$ = 0 and $k_y$ = 0 planes) is protected by the mirror symmetry and TRS. 

To visualize the predicted KNL along $\Gamma-Z$, we carried out detailed photon-energy dependent ARPES investigations in both SX and VUV regimes on the (001) cleaving surface (Fig. S5 in Supplementary Note 3). SX measurements provide better $k_z$ resolution while VUV results in better energy and in-plane momentum resolution~\cite{Sobota2021}. According to our DFT calculations (Fig.~\ref{fig:Basic}d), the KNL along $\Gamma-Z$ closest to the Fermi level on the occupied side extends from -1.5 to -0.85 eV, and forms the band bottom of an electron band in the in-plane direction. A direct visualization of the $\Gamma-Z$ KNLs is displayed in Fig.~\ref{fig:KNL}b, where surface states and $k_z$ broadening effects are well mitigated in the SX regime (see Fig. S5). DFT calculated bulk bands indicating the KNLs superimposed on the image (Fig.~\ref{fig:KNL}b) are in good agreements with SX ARPES data.
We can also examine the KNL from measurements in the VUV regime. Specifically, from the in-plane directions along both $\Gamma-\Sigma$ and $Z-\Sigma_1$ we observe the KNL connecting to the electron band bottom, consistent with the prediction of the DFT calculations. The evolution of the KNL along $k_z$ can be better illustrated by the shift of the electron band bottom from a series of in-plane dispersions at different photon energies, where 114 eV and 132 eV approximately correspond to the $k_z$ = 2$\gpi/c$ and $k_z$ = 0 planes, respectively (Fig.~\ref{fig:KNL}d). Despite surface states and $k_z$ broadening effect in the VUV regime (Fig. S5), the spectral intensity shift from higher binding energy (at $Z$) to lower binding energy (near $\Gamma$) is evident in the
energy distribution curves (EDCs) at the $\bar{\Gamma}$ point taken from $k_z$ = 0, $k_z$ = $\gpi/c$, and $k_z$ = 2$\gpi/c$ where the peak in the EDC corresponds to the calculated electron band bottom in the overlaid DFT bulk calculation (Fig.~\ref{fig:KNL}f), confirming the bulk KNLs along $k_z$.

As both the theory and the data indicate the presence of KNLs protected by mirror symmetries through the combined studies of DFT and ARPES methods, we demonstrate the double degeneracy of the KNL from the in-plane direction. As seen from the DFT calculations (Fig.~\ref{fig:Basic}d), the bands connected to the KNLs are split due to SOC in the in-plane direction. Experimentally, we observe a band splitting along $\bar{\Gamma}-\bar{X}$ direction (Fig.~\ref{fig:KNL}c) in the band emanating from $\bar{\Gamma}$. The splitting can be well resolved from a zoomed-in view marked by the red box, where the momentum distribution curve (MDC) shows clearly a double-peak structure that can be fitted by two Lorentzians on a constant background. From the comparison with an ab-initio surface calculation for the corresponding $\Gamma-X$ cut in Fig.~\ref{fig:KNL}e, the observed band splitting should predominantly come from surface states. Nonetheless, from Fig.~\ref{fig:KNL}e and Fig.~\ref{fig:Basic}d, especially clear is the splitting of all bands away from $\bar{\Gamma}$, regardless of their bulk or surface nature due to broken IS and preseved TRS, consistent with the 2-fold nature of the KNL along $\Gamma-Z$.

\subsection{Weyl fermions}

Besides KNLs, the lack of IS 
in SmAlSi also results in a system of WFs. WFs in both the paramagnetic and magnetic states have already been reported in isostructural compounds $R$AlSi/$R$AlGe ~\cite{Chang2018,Xu2017,Suzuki2019,Meng2019,Sanchez2020,Destraz2020,Liu2021,Sun2021,Gaudet2021,Wang2022_1,Wang2022_2,Sakhya2022,Piva2021,Xu2021,Cao2022,Xu2021_squid1,Xu2021_squid2,Zhao2022,Yang2021}. In SmAlSi, we identify a total of 56 WFs classified into 5 categories, W1, W2, W3, W3' and W4 based on their locations in the BZ (Fig.~\ref{fig:WF}a and b for the top and side views). As shown in Fig.~\ref{fig:WF}b and Table~\ref{tab1}, we find three types of WFs (W1, W3, and W3') that are pinned to the $k_z$ = 0 plane by the $C_{2z}\hat{\mathcal{T}}$ symmetry (solid circles), while W2 and W4 WFs are symmetrically distributed on the two sides of the $k_z$ = 0 plane (empty circles). To show the Weyl fermion band dispersions, we take 6 cuts (see Fig.~\ref{fig:WF}a for positions) measured in the VUV regime, overlaid by the DFT bulk calculations (Fig.~\ref{fig:WF}e-f). The corresponding ab-initio surface calculations are also shown in Fig.~\ref{fig:WF}g-h. We discuss the detailed band comparison and assigment in the Supplementary Note 4. Nonetheless, the consistency between calculation and ARPES observation is evident. For instance, Cut 1 shows the linear bulk dispersion below W1 and its parabolic behavior at deeper binding energies. Cut 3 shows mainly three hole-like band features well captured by the ab-initio calculation projected onto the surface, with the middle one forming W2. In particular, we find the best match between DFT and ARPES data near W1 and W2 after shifting the DFT bands up by 33 meV, which is determined by our quantum oscillations experiments described below. For bands in the W3, W3' and W4 region, however, an upward shift of 120 meV is needed for both bulk (magenta lines) and surface calculations to capture the two hole-like bands in Cut 4 and 5, as well as the W-like dispersion in the middle part of Cut 6 (see Fig.~\ref{fig:WF}f and h). Crucially, we find that the experimentally observed upper hole-like bands in Cut 4 and Cut 5 are related with the Fermi arc states, while the lower hole bands correspond to the bulk bands below WFs pointed out in Cut 4, 5, and 6 of Fig.~\ref{fig:WF}f. We emphasize that the observed VUV spectra here (Fig.~\ref{fig:WF}e-f) have strong surface contribution. Therefore, band assignment is not a straightforward task, and we discuss it in detail in Supplementary Note 4. To further understand the band structure near the W3'-W3-W4 region, Fig.~\ref{fig:WF}c shows a measured FS in the second quadrant where a dumbbell-like pattern with two tilted ends is observed. The intensities on the two ends mainly consist of the Fermi arc states (upper bands in Cut 4 and Cut 5 in Fig.~\ref{fig:WF}f) at the Fermi level connecting WFs, while the bar in the middle of the dumbbell comes from the bulk bands below WFs (hole band top below W3 at E$_{\rm F}$ in Cut 6 and lower bands in Cut 4 and 5 of Fig.~\ref{fig:WF}f). Their band dispersions are further investigated by two loop cuts (red loop 1 and blue loop 2) taken on Fig.~\ref{fig:WF}c. Such method has been implemented to distinguish Fermi arcs from normal bands~\cite{Sanchez2020,Sakhya2022}. The results are displayed in Fig.~\ref{fig:WF}d. Clearly, one can see a two-band configuration here (green dashed guideline for the Fermi arc band and magenta dashed guideline for the bulk band in Fig.~\ref{fig:WF}d1 and Fig.~\ref{fig:WF}d3). The upper green band shows a right-moving and a left-moving chiral mode in Loop 1 and Loop 2, respectively, further supporting its Fermi arc origin. The lower magenta band, identified as the bulk band below WFs contributing to the bar of the dumbbell, connects back to itself after taking a full Loop 1 or 2, despite that its intensities are suppressed at the boundary of the loop likely due to the matrix element effects.

Since VUV ARPES is more surface sensitive and subject to $k_z$ broadening of bulk bands, as exemplified by the comparison in Fig.~\ref{fig:WF}e-h, we further conducted SX ARPES to confirm the bulk nature of the WFs. Figure~\ref{fig:WF}i-l shows the SX data for W1, W2, W3' and W3, superimposed by bulk calculations (magenta lines), where the bulk nature of those WFs is confirmed by the excellent agreement between calculations and experiment. Moreover, to visualize the Fermi arcs directly connected to W3, W3', and W4 WFs above the Fermi level, we show the calculated DFT surface constant energy contours (Fig.~\ref{fig:WF}m1-m3) at the respective Weyl fermion energy positions (Table~\ref{tab1}). Fermi arcs can be clearly seen in each case.

\begin{figure}[t]
    \centering
    \includegraphics[width=450pt]{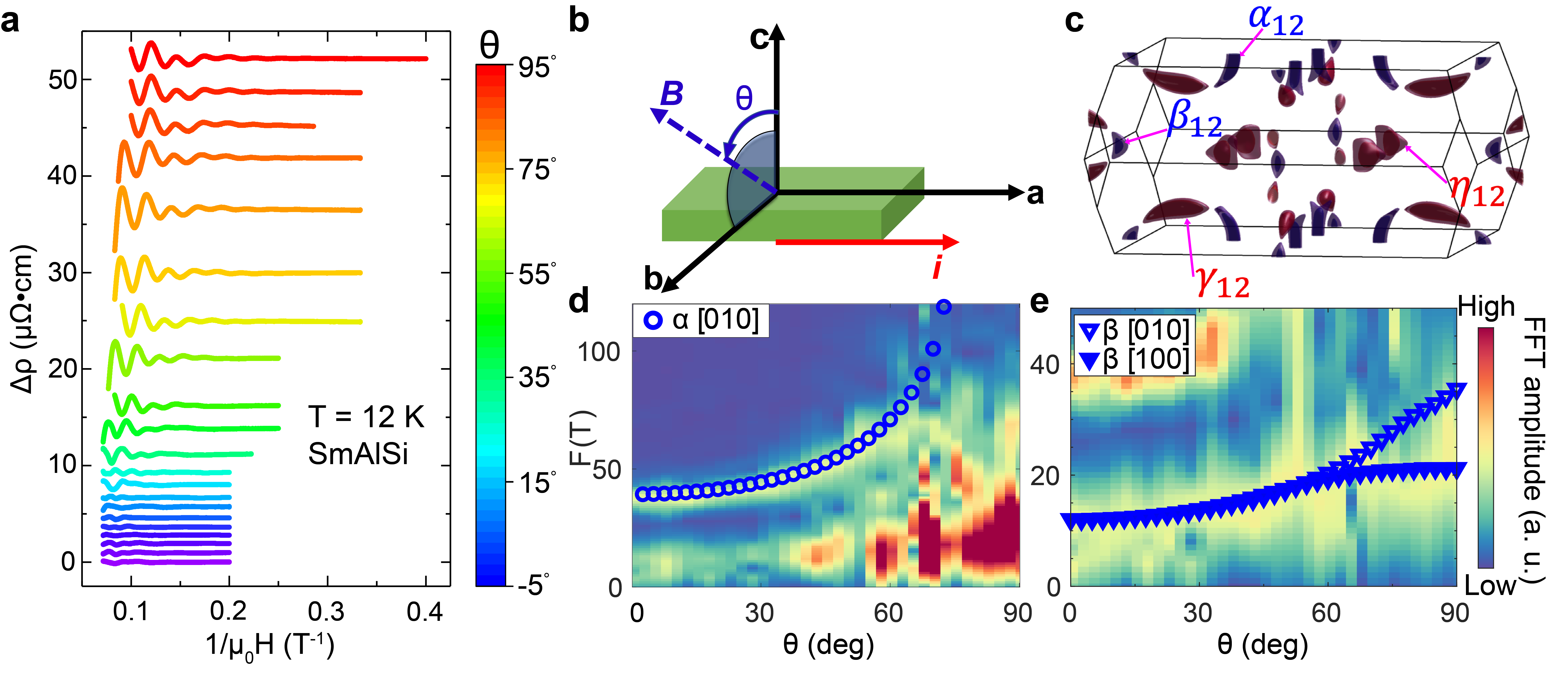}
    \caption{\textbf{Angle dependent quantum oscillation measurements of SmAlSi at 12 K. } \textbf{a} Waterfall plot of the Shubnikov-de Haas oscillations after background subtraction. A vertical offset is applied for better visualization. \textbf{b} Schematics of the experimental setup for (\textbf{a}). \textbf{c} Fermi surface from first-principles calculations showing eight distinct Fermi pockets located at four positions in the BZ. The electron (hole) pockets are denoted by the red (blue) color. \textbf{d,e} Contour plot of the Fast Fourier Transform spectrum from (\textbf{a}), highlighting the components of $F>$35 T and $F<$30 T, respectively. First-principles estimated frequencies of pocket $\alpha$ (cricles in (\textbf{d})) and pocket $\beta$ (triangles in (\textbf{e})). The [010] (open symbols) and [100] (closed symbols) represent the Fermi pocket along [100] direction and [010] of the same type in the BZ, respectively. }
    \label{fig:QO}
\end{figure}

In addition to ARPES, we have also performed angle-dependent quantum oscillations (QOs) measurements to corroborate the understanding of the FS and the underlying topology. The resistivity measurements are performed in the paramagnetic state (at T = 12 K $>$ T$_{\rm N}$), with the magnetic field oriented in the $bc$ plane and the current applied along the $a$ axis (Fig.~\ref{fig:QO}b). After a smooth background subtraction, strong Shubnikov-de Haas oscillations are observed for all measured field orientations (Fig.~\ref{fig:QO}a). The oscillation frequencies are obtained from Fast Fourier Transforms of the measured data  (Fig.~\ref{fig:QO}a), and are plotted as contour maps in Fig.~\ref{fig:QO}d-e. We further compare the measured oscillation frequencies with Fermi surface cross sections from DFT calculations, where the proportionality between quantum oscillation frequencies $F$ and extremal Fermi surface $A_{\rm ext}$ is given by the Onsager relationship: 
\begin{equation}
F=\frac{\hbar A_{\rm ext}}{2\gpi e},
\end{equation}
where $\hbar$ is the Planck constant and $e$ is the electron charge. The DFT calculations with SOC (Fig.~\ref{fig:Basic}d) indicate four bands crossing the Fermi energy, each forming two small Fermi pockets with distinct shapes in the first BZ.  We notice that the Fermi pocket locations in the DFT calculation with SOC are similar to those without SOC as previously reported~\cite{Cao2022}. As a result of the broken IS and SOC, the band degeneracy is lifted and the number of pockets is doubled. These pockets are denoted as $\alpha_{1,2}$, $\beta_{1,2}$, $\gamma_{1,2}$, $\eta_{1,2}$ with the subscript 1 and 2 representing the pair of SOC-split bands, as shown in Fig.~\ref{fig:QO}c. We identify the QOs from pockets $\alpha_1$ and $\beta_1$ in the Shubnikov-de Haas signal, as the trend of extremal Fermi surface cross section from DFT (symbols in Fig.~\ref{fig:QO}d,e) matches well with observed values after an upshift of 33 meV. Given that those pockets and corresponding bands are responsible for the W1 Weyl points and W2 Weyl points, our results provide strong evidence that those Weyl points are located close to the Fermi energy as predicted in the DFT and observed by ARPES. 

We now consider the relation between the magnetic order in SmAlSi (Fig.~\ref{fig:Basic}f-h) and the electronic properties embedded in the FS determined experimentally (ARPES and QOs) and theoretically (DFT). In the isostructural compound NdAlSi, it has recently been proposed that Weyl pockets could give rise to nesting-induced incommensurate SDW~\cite{Gaudet2021}. To evaluate this possibility in the case of SmAlSi, we note that our experimentally-observed FS shows a flat dispersion that could potentially provide nesting conditions along the (110) direction (see Fig.~\ref{fig:WF}c and the \textbf{q}$_{\rm N}$ denoted in Fig.~\ref{fig:nesting}a), where the incommensurate magnetic wavevector was observed below T$_{\rm N}$ = 11.5 K~\cite{Yao2022}. Importantly, the nesting feature consists of the lower hole-like bands emphasized in Fig.~\ref{fig:WF}d (magenta dashed lines) and Fig.~\ref{fig:WF}f, rather than any Weyl fermion bands. Further assessing the nesting condidtion quantitatively, five cuts perpendicular to the nesting feature are extracted for momentum distribution curve (MDC) analysis (Fig.~\ref{fig:nesting}b), where we fit the peak locations in each MDC taken from E$_{\rm F}$ to extract the location of the Fermi surface (see details in Fig. S7). The fitted peak values are subsequently plotted in Fig.~\ref{fig:nesting}a,b. From the average of the fitted positions (see Fig.~\ref{fig:nesting}d), we find a wavevector of \textbf{q}$_{\rm N}$ = (0.74,0.74,0) that connects the experimentally observed flat features across the BZ, as schematically illustrated in Fig.~\ref{fig:nesting}a (see Supplementary Note 5 for details). Such nesting wavevector is also corroborated by the auto-correlation analysis presented in the Supplementary Information Fig. S9. We do note that the incommensurability is likely sensitive to sample-dependent variations of the chemical potential, and careful comparison of this with precise determination of the magnetic wavevector is important to establish the mechanism of nesting-driven magnetic order in this material. To match the wavevector observed by neutron scattering~\cite{Yao2022}, a large shift of 100 meV of the chemical potential would be needed (see Fig. S8). Furthermore, to check for nesting instability, 
we carried out Lindhard susceptibility calculations after adjusting the chemical potential shift between experimental and DFT data. The bulk FS is shown in Fig.~\ref{fig:nesting}c with nesting vectors of interest indicated by \textbf{q}$_{\rm 1}$, \textbf{q}$_{\rm 2}$, and \textbf{q}$_{\rm 3}$. Consequently, the imaginary part of the Lindhard function peaks at \textbf{q}$_{\rm 1}$ = (0.44$\pm$0.04, 0.44$\pm$0.04, 0) and \textbf{q}$_{\rm 2}$ = (0.70$\pm$0.04, 0.70$\pm$0.04, 0) (Fig.~\ref{fig:nesting}e). However, no peaks are found in the real part of the Lindhard function (Fig.~\ref{fig:nesting}f), hence casting further doubt on a nesting-driven scenario~\cite{Johannes2008,Inosov2009}. Here $\pm$0.04 represents the range of sub-peaks upon the main nesting vectors which is due to the various nesting possibilities for the two Weyl pockets with finite sizes. The unit of the nesting vectors is set as half of the BZ length, in order to be consistent with previous studies on NdAlSi~\cite{Gaudet2021} and SmAlSi~\cite{Yao2022}. To recover the conventional unit which is the length of the reciprocal vectors, an extra ratio 1/2 should be multiplied to the original \textbf{q}$_{\rm 1,2}$, as well as the observed \textbf{q} in ARPES. In addition, we note that the Weyl fermions in SmAlSi originate from the Sm atomic orbitals instead of the electron spin (Fig. S4), which can affect the form of the Weyl-nesting-induced RKKY interaction and may lead to novel magnetic orders.


\section*{Discussion and Conclusion} 
Using both experimental measurements (ARPES and QOs) and theoretical calculations (DFT), we have presented evidence suggesting the presence of KNLs hosted by the paramagnetic phase of SmAlSi, located away from the Fermi level. Therefore, further investigation, both theoretical and experimental, to disambiguate the presence of KNLs and optimize them for the appearance of topological responses in transport or optics experiments are crucial. It is important to note that KNLs are distinct from the nodal lines resulting from band inversion. The former are robust against SOC and protected by the combination of the TRS and achiral crystal symmetries while the latter are not. This mechanism also offers an avenue for tunability such that, when the mirror or roto-inversion symmetries are broken, the KNL metal can be transformed into Kramers Weyl semimetal. Regarding the magnetism that emerges in SmAlSi, our evaluation of the possible nesting conditions using measured and calculated FS reveals that the potential nesting wavevectors deviate from the incommensurate SDW wavevector near 1/3 or 2/3 measured by neutron experiments~\cite{Yao2022}, and the real part of the Lindhard function does not show any prominent peak structure (See Fig.~\ref{fig:nesting}f). Note that it is meaningful to compare the band structure of SmAlSi in its paramagnetic state with its lower-temperature magnetic wavevector, since the material spontaneously condenses to the magnetic state without external field. In principle, fully spin polarized state is a result of field-driven crossover but not regarded as an onset of any instabilities. Furthermore, the low-temperature ARPES data summarized in Supplementary Note 6 Fig. S10 and Fig. S11 do not show band reconstruction related to the magnetic phase transition, consistent with a recent report~\cite{Rui2023}. Therefore, it is unlikely that the SDW is purely driven by a Weyl pocket nesting instability. Considering the broken IS in this material, Dzyaloshinskii-Moriya interaction could play an important role in stabilizing the spiral order, while the RKKY interaction should be mediated by both conventional~\cite{Zhentao2020} and Weyl electrons. The discrepancy between the magnetic wavevector and FS nesting conditions might come from higher-order spin-spin interactions. Further studies on such a mechanism are required to account for why certain magnetic wavevectors eventually become energetically favorable. Given the range of exotic magnetic structures that develop at low temperatures in the LaPtSi type family of compounds~\cite{Chang2018,Suzuki2019,Meng2019,Sanchez2020,Destraz2020,Liu2021,Sun2021,Gaudet2021,Wang2022_1,Wang2022_2,Sakhya2022,Piva2021,Xu2021,Cao2022,Yao2022,Zhao2022,Xu2021_squid1,Xu2021_squid2,Yang2021}, this family of materials provides an ideal platform to study novel band topology and its interplay with many-body phenomena especially regarding the spin degree of freedom. 

\section*{Acknowledgments}
We thank Zhentao Wang, Gregory Mccandless and Julia Chan for useful discussions. This work was mainly supported by the Department of Defense, Air Force Office of Scientific Research under Grant No. FA9550-21-1-0343. This research used resources of the Stanford Synchrotron Radiation Lightsource, SLAC National Accelerator Laboratory, which is supported by the U.S. Department Of Energy (DOE), Office of Science, Office of Basic Energy Sciences under Contract No. DE-AC02-76SF00515. Part of the research described in this work was also performed at the Canadian Light Source, a national research facility of the University of Saskatchewan, which is supported by Canada Foundation for Innovation (CFI), the Natural Sciences and Engineering Research Council of Canada (NSERC), the National Research Council (NRC), the Canadian Institutes of Health Research (CIHR), the Government of Saskatchewan, and the University of Saskatchewan. 
The ARPES work at Rice University was also supported by the Gordon and Betty Moore Foundation's EPiQS Initiative through grant No. GBMF9470. EM acknowledges partial support from the the Robert A. Welch Foundation grant C-2114. K.T.L. acknowledges the support of HKRGC through RFS2021-6S03, C6025-19G, AoE/P-701/20, 16307622, 16310520 and 16310219. Work at University of California, Berkeley, is funded by the U.S. Department of Energy, Office of Science, Office of Basic Energy Sciences, Materials Sciences and Engineering Division under Contract No. DE-AC02-05-CH11231 (Quantum Materials program KC2202). The SHG experiment is sponsored by the Vagelos Institute of Energy Science and Technology graduate fellowship and the Dissertation Completion Fellowship at the University of Pennsylvania. L.W. acknowledges the support by the Air Force Office of Scientific Research under award number FA9550-22-1-0410. 

\section*{Methods}
\subsection{Crystal growth.}

Single crystals of SmAlSi were grown by a self-flux method. Sm, Al and Si were used as starting elements and were mixed with composition Sm:Al:Si=1:10:1 in an alumina crucible, and sealed up in an evacuated quartz tube. The tube was heated up to 1000$\degree$C in 6 hours, held for 12 hours and the excess flux decanted at 700$\degree$C after a cooling at 0.1$\degree$C min$^{\rm -1}$. 

\subsection{Transport and magnetization measurements.}

Transport and magnetization measurements were performed in a Quantum Design Dynacool Physical Property Measurement System. The electrical resistivity was measured with a standard four-probe technique and the heat capacity was measured with a relaxation method. The magnetization is measured using the vibrating sample magnetometer (VSM) option.

\subsection{Second harmonic generation.}

The SHG measurement is performed at room temperature. We use a Ti-Sapphire laser (30 fs, 80 MHz) with a central wavelength of 800 nm. The laser is focused on the (001) facet under 45 degree incidence by a 10X objective. The beam diameter is estimated to be 10 $\rm\mu$m. The SHG light is collected by a lens, and then into a photo-multiplier tube. The polarizations of the incident light and the detecting light are simultaneously controlled by a half-wave plate and a linear polarizer respectively.

\subsection{ARPES measurements.}

Angle-resolved photoemission spectroscopy (ARPES) measurements in the vacuum ultraviolet (VUV) regime were performed at the Advanced Light Source, Beamline 4.0.3 (MERLIN), equipped with a SCIENTA R8000 analyzer, and at Stanford Synchrotron Radiation Lightsource (SSRL), Beamline 5-2, equipped with a DA30 electron analyzer. The single crystals were cleaved in-situ at 13 K (above any magnetic transition temperature of SmAlSi) and measured in ultra-high vacuum with a base pressure better than $5\times10^{-11}$ Torr. Photon energy-dependent measurements were carried out from 30 to 180 eV with. Energy and angular resolution were set to be better than 10 meV and 0.1$^{\circ}$, respectively, for data obtained below 60 eV. Soft X-ray (SX) ARPES measurements were performed at the QMSC beamline of the Canadian Light Source, equipped with a R4000 electron analyzer. SmAlSi samples were cleaved \textit{in-situ} and measured under a vacuum condition better than $6\times10^{-11}$ Torr and temperature stabilized at 17 K. The energy and angular resolution of the SX measurements are around 120 meV and 0.3$^{\circ}$, respectively. All the data shown are taken using linear horizontal (LH) light polarization. The VUV experimental setups use vertical analyzer slit while the SX experimental setup uses a horizontal analyzer slit. Therefore the LH polarization for the VUV (SX) ARPES experimental geometry is perpendicular (parallel) to the cut direction.

\subsection{DFT calculations.}

We used the Vienna Ab initio Simulation Package (VASP)~\cite{DFT1} with the Perdew-Berke-Ernzerhof's
(PBE) exchange-correlation functional in the generalized-gradient
approximation~\cite{DFT2,DFT3} to perform the density functional theory (DFT) calculations~\cite{DFT4}. We adopted the experimental lattice constants and crystal structure (shown as the cif file of Supplementary Data 1).
As we focused on the nonmagnetic phase, the 4f electrons
of Sm atoms were kept in the core. Both cases without and with spin-orbit coupling (SOC) were calculated, and a Wannier tight-binding model was further obtained through the Wannier90 package \cite{WANN}, which accurately fits the DFT bands with the inclusion of SOC. From this Wannier model, we studied in detail all the topological crossings including the KNLs, Weyl nodes and the hourglass nodal lines in the nonmagnetic SmAlSi.
To achieve the best match between ARPES data and DFT calculation of the WFs, we used a 0.15 eV surface potential for first-principles surface calculation (See Supplementary Information Fig. S3).

\section*{Data availability}
The data that support the findings of this study are available from the corresponding authors upon reasonable request.

\section*{Author contributions} The project was organized by MY, EM, and KTL. The sample synthesis and characterization was carried out by YG, SL, and EM. The ARPES work was done by YZ, JO, JH, ZY, MY, and RJB, with the help of JD, MH, DL, MZ, and SG. The SHG was carried out by ZN and LW. The first-principles calculations were carried out by XG and KTL. YZ, YG, and XG contributed equally to this work. All authors contributed to manuscript preparation.

\section*{Competing Interests} The authors declare no
competing interests. Professor Ming Yi is an Editorial Board Member for Communications Physics, but was not involved in the editorial review of, or the decision to publish, this article.

\newpage

\begin{table}
\centering{}%
\begin{tabular}{cccccc}
\hline 
Weyl nodes \hspace{0.2cm} & Position: $(k_{x},k_{y},k_{z})(1/\text{Å})$\hspace{0.2cm} & Chirality\hspace{0.2cm} & $E_{\rm DFT}$ (meV)\hspace{0.2cm} & $E_{\rm shifted}$ (meV)\hspace{0.2cm} & Number\tabularnewline
\hline 
W1 & $(0.783,0.004,0)$ & $+1$ & $87$ & $120$ & 8\tabularnewline
W2 & $(0.374,0.032,0.290)$ & $-1$ & $16$ & $49$ & 16\tabularnewline
W3 & $(0.312,0.264,0)$ & $-1$ & $-78$ & $42$ & 8\tabularnewline
W3' & $(0.329,0.290,0)$ & $-1$ & $-71$ & $49$ & 8\tabularnewline
W4 & $(0.302,0.271,0.035)$ & $+1$ & $-85$ & $35$ & 16\tabularnewline
\hline 
\end{tabular}\caption{\textbf{Five groups of Weyl nodes in nonmagnetic SmAlSi near the Fermi energy
E$_{\rm F}$}. We only list one Weyl node for each group within one sixteenth
of the first Brillouin zone. Other Weyl nodes that are not listed
in the table can be derived using time-reversal
symmetry and the crystalline symmetry operations. W1, W3 and W3'
Weyl nodes are pinned to the $k_z$ = 0 plane by the $C_{2z}\hat{\mathcal{T}}$
symmetry. $E_{\rm DFT}$: Energy positions of the Weyl nodes from first-principles calculation. $E_{\rm shifted}$: Energy positions of the Weyl nodes after shifting the Fermi level as discussed in the text.\label{tab1}} 
\end{table}

\begin{figure}[t]
    \includegraphics[width=470pt]{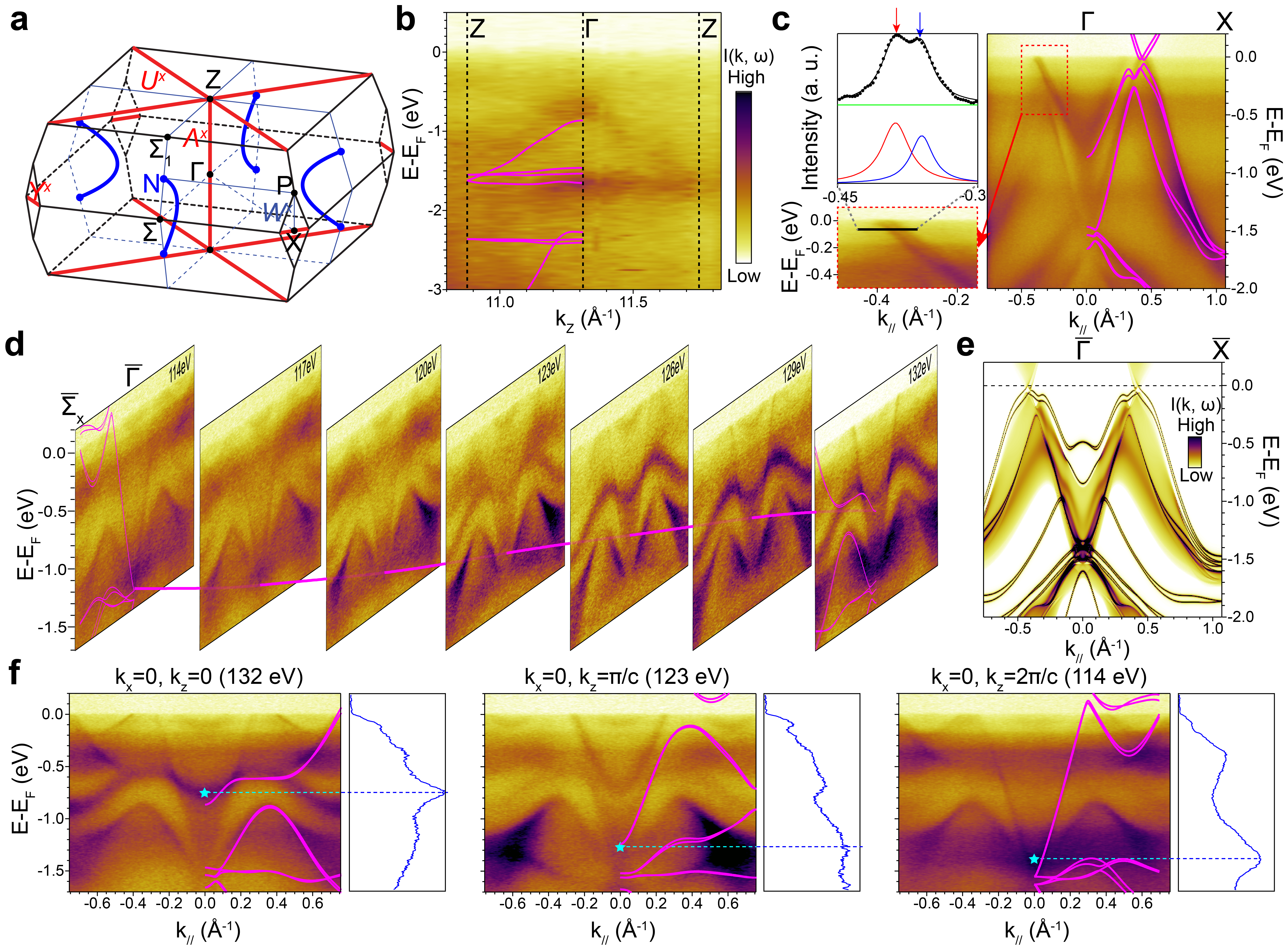}
    \centering
    \caption{\textbf{Kramers nodal lines.} \textbf{a} Illustration of different types of KNLs where red (blue) lines connect different (same) TRIM points. \textbf{b} Measured dispersions along the $\Gamma-Z$ direction by soft X-ray photons (shown from 458 eV to 546 eV) overlaid with bulk calculations. \textbf{c} Spectral image along $\Gamma-X$ (at 63 eV) to show band splitting away from the KNL, overlaid with calculated bulk bands. MDC integrated between -60 and -70 meV is fitted with two Lorentzian peaks and a constant background to resolve the band splitting.
    \textbf{d} Photon energy scan showing the spectral weight transfer from near $k_z$ = 2$\gpi/c$ (114eV) to near $k_z$ = 0 (132 eV) for the $\bar{\Gamma}-\bar{\Sigma_x}$ counterparts,  where the magenta line threading through cuts is the calculated KNL corresponding to the highest band in}
    \label{fig:KNL}
\end{figure}

\makeatletter
\setlength\@fptop{0pt} 
\setlength\@fpsep{8pt plus 1fil} 
\setlength\@fpbot{0pt}
\makeatother

\begin{figure}[t!]
  \contcaption{(continued caption) panel (\textbf{b}). \textbf{e} Ab-initio surface calculation for the $\Gamma-X$ cut. \textbf{f} Evolution of energy distribution curve of the $k_x$ = 0 cuts at different $k_z$'s, integrated from $k_y$ = -0.05 $\angstrom^{-1}$ to 0.05 $\angstrom^{-1}$, overlaid with bulk calculations. Horizontal dotted lines and cyan stars show the peak positions corresponding to the $\Gamma-Z$ KNL intensity.}
\end{figure}

\begin{figure}[t]
    \includegraphics[width=470pt]{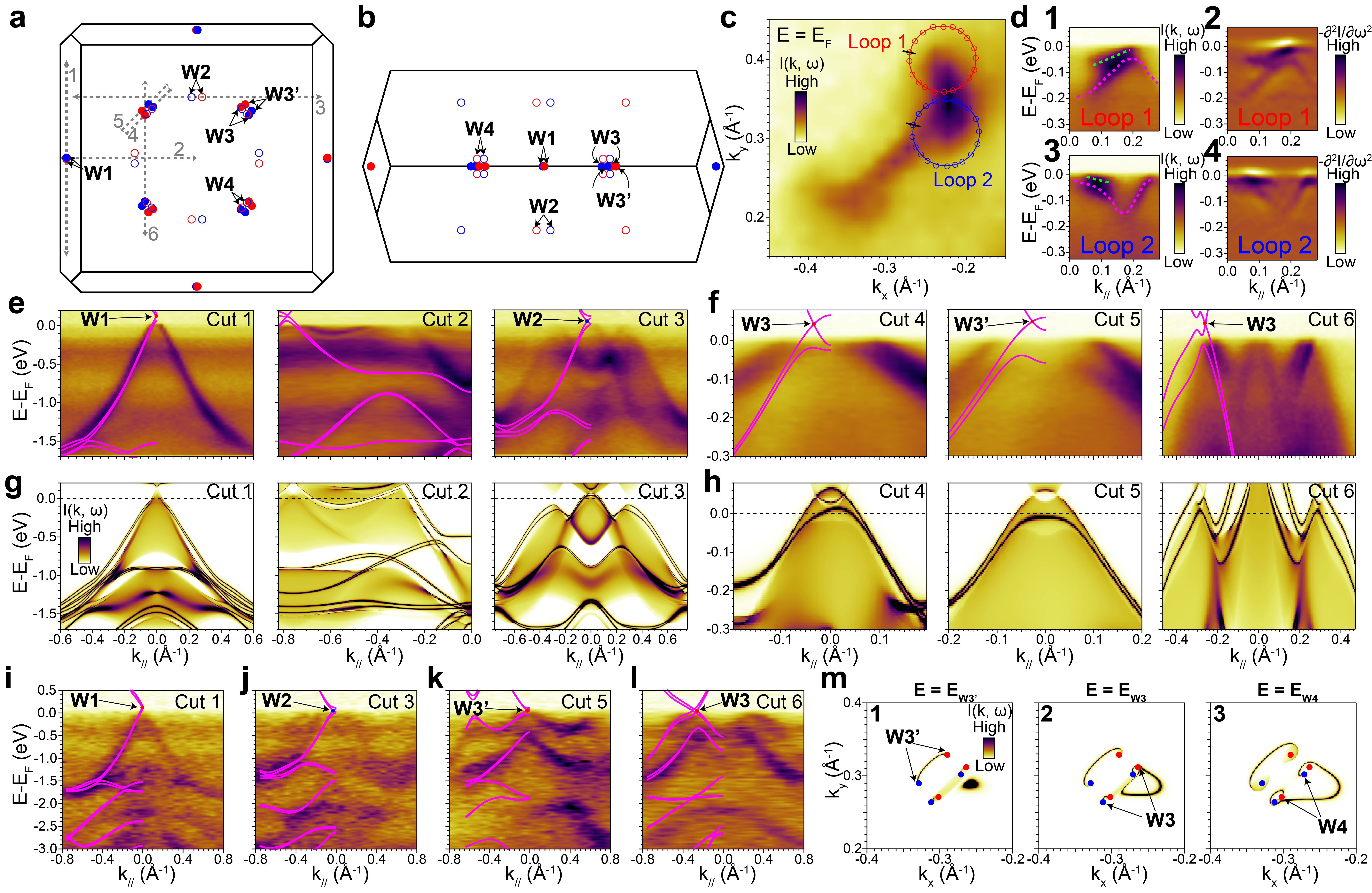}
    \centering
    \caption{\textbf{Weyl fermions and Fermi arcs.} \textbf{a,b} Locations of all Weyl fermions from the top view and side view of the BZ. \textbf{c} ARPES measured Fermi surface (FS) focusing on the W3-W3'-W4 region at 44.5 eV. Loop 1 and Loop 2 are represented by the red and blue circles, respectively. Red and blue markers on the loops indicate actual data points taken before interpolation. Black lines on the loops represent starting and ending positions. The loops are taken clockwise. \textbf{d} Loop cuts taken from (\textbf{c}). (\textbf{d1}) and (\textbf{d3}) are the interpolated data for Loop 1 and Loop 2, respectively. (\textbf{d2}) and (\textbf{d4}) are the corresponding second derivative plots. Green and magenta dashed lines in (\textbf{d1}) and (\textbf{d3}) serve as eye guidance for the upper Fermi arc band and the lower bulk band. \textbf{e}, \textbf{f} Band dispersions measured in the VUV regime (63eV for (\textbf{e}) and 44.5 eV for (\textbf{f})) related to Weyl fermions W1, W2, W3 and W3' denoted as Cut 1 to 6 in (\textbf{a}), with DFT bulk calculations overlaid. \textbf{g}, \textbf{h} Corresponding}
    \label{fig:WF}
\end{figure}

\makeatletter
\setlength\@fptop{0pt} 
\setlength\@fpsep{8pt plus 1fil} 
\setlength\@fpbot{0pt}
\makeatother

\begin{figure}[t!]
  \contcaption{(continued caption) ab-initio surface calculations for (\textbf{e}) and (\textbf{f}), respectively. \textbf{i}-\textbf{l} Band dispersions for W1, W2, W3', and W3 measured by soft X-rays at 498 eV. (\textbf{j}) is taken at the 2nd BZ of the 498 eV cut which actually crosses the W2 $k_z$ plane. \textbf{m} Ab-initio surface calculation of constant energy contours taken at the energies of W3', W3, and W4 to show the Fermi arc structure. DFT calculations in the region of W1 and W2 are shifted up by 33 meV, and those calculated in the region of W3 and W3' are shifted up by 120meV. All the ab-initio surface calculations implement $V_{\rm surf}$=0.15 eV.}
\end{figure}

\clearpage

\begin{figure}[t]
    \includegraphics[width=370pt]{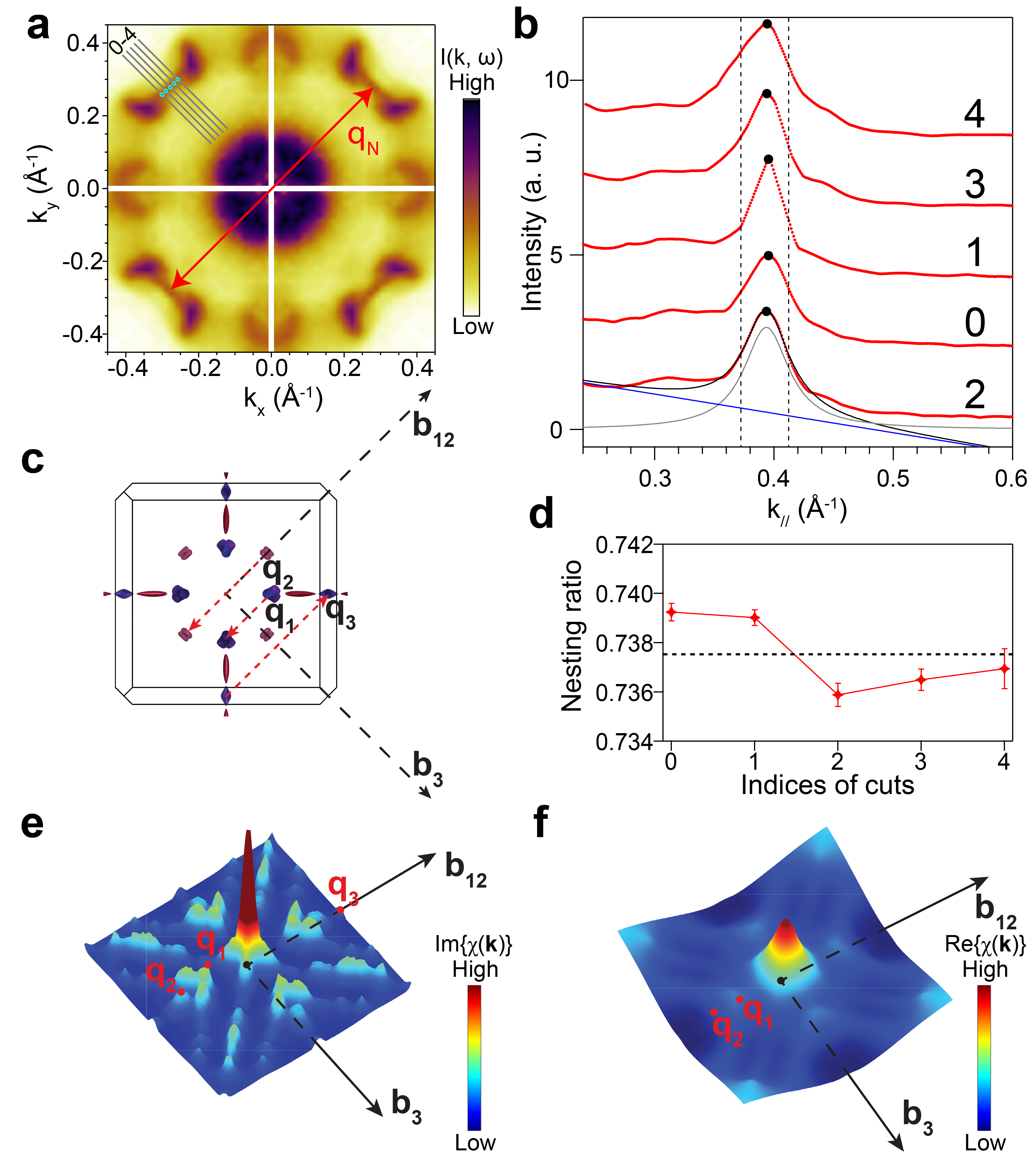}
    \centering
    \caption{\textbf{Evaluation of nesting possibilities on the Fermi surface.} \textbf{a} $C_4$ symmetrized FS with gray lines along $(\bar{1}10)$ indicating the 5-cut directions. Open cyan circles are the fitted band positions. Red arrow denotes the nesting vector \textbf{q}$_{\rm N}$. \textbf{b} MDCs at E$_{\rm F}$ with 3 meV integration window for Cuts 0 - 4. Vertical offsets are applied to the five MDCs for clear visualization. Two vertical black dashed lines indicate the fitting region. The fitting result of Cut 2 from a gray Lorentzian peak and a blue}
    \label{fig:nesting}
\end{figure}

\makeatletter
\setlength\@fptop{0pt} 
\setlength\@fpsep{8pt plus 1fil} 
\setlength\@fpbot{0pt}
\makeatother

\begin{figure}[t!]
  \contcaption{(continued caption) linear background is plotted as the black curve at the bottom. The solid black dots show the fitted peak positions for Cut 0 - 4. \textbf{c} The top view of the electron (red) and hole (blue) pockets within the first BZ of paramagnetic SmAlSi at the experimental chemical potential. $\bm{b}_{12}$ and $\bm{b}_3$ are two reciprocal vectors lying within the $k_z$ = 0 plane. \textbf{d} Nesting ratios extracted from Cuts 0 - 4 with error bars included, where the horizontal black dashed line indicates the arithmetic mean value of the five nesting ratios. Errors bars are first calculated from the standard error of the fitting and converted following error propagation formulae. \textbf{e} and \textbf{f} are respectively the imaginary and real part of the Lindhard function in the static limit $\omega\rightarrow0$ and within the $k_z=0$ plane. The potential nesting peaks in (\textbf{e}) and the interaction instability vectors in (\textbf{f}) correspond to the vectors in (\textbf{c}) which connect different Weyl Fermi pockets.}
\end{figure}

\end{document}